\documentclass{PoS}
\usepackage{graphicx}
\usepackage{epsfig}

\def\kevc1{\ifmmode\mathrm{\ keV/{\mit c}}
          \else$\mathrm{\ keV/{\mit c}}$\fi}
\def\Mevc1{\ifmmode\mathrm{\ MeV/{\mit c}}
          \else$\mathrm{\ MeV/{\mit c}}$\fi}
\def\mevc1{\ifmmode\mathrm{\ MeV/{\mit c}}
          \else$\mathrm{\ MeV/{\mit c}}$\fi}
\def\gevc1{\ifmmode\mathrm{\ GeV/{\mit c}}
          \else$\mathrm{\ GeV/{\mit c}}$\fi}
\def\kevc2{\ifmmode\mathrm{\ keV/{\mit c}^2}
          \else$\mathrm{\ keV/{\mit c}^2}$\fi}
\def\Mevc2{\ifmmode\mathrm{\ MeV/{\mit c}^2}
          \else$\mathrm{\ MeV/{\mit c}^2}$\fi}
\def\Gevc2{\ifmmode\mathrm{\ GeV/{\mit c}^2}
          \else$\mathrm{\ GeV/{\mit c}^2}$\fi}
\def\Gev2c2{\ifmmode\mathrm{\ GeV^2/{\mit c}^2}
          \else$\mathrm{\ GeV^2/{\mit c}^2}$\fi}

\def\ubar{\ifmmode\mathrm{\overline {u}}
          \else$\mathrm{\overline{u}}$\fi}
\def\dbar{\ifmmode\mathrm{\overline {d}}
          \else$\mathrm{\overline{d}}$\fi}
\def\sbar{\ifmmode\mathrm{\overline {s}}
          \else$\mathrm{\overline{s}}$\fi}
\def\cbar{\ifmmode\mathrm{\overline {c}}
          \else$\mathrm{\overline{c}}$\fi}
\def\bbar{\ifmmode\mathrm{\overline {b}}
          \else$\mathrm{\overline{b}}$\fi}
\def\tbar{\ifmmode\mathrm{\overline {t}}
          \else$\mathrm{\overline{t}}$\fi}
\def\qbar{\ifmmode\mathrm{\overline {q}}
          \else$\mathrm{\overline{q}}$\fi}

\def\uq{\ifmmode\mathrm{u}
          \else$\mathrm{u}$\fi}
\def\dq{\ifmmode\mathrm{d}
          \else$\mathrm{d}$\fi}
\def\sq{\ifmmode\mathrm{s}
          \else$\mathrm{s}$\fi}
\def\cq{\ifmmode\mathrm{c}
          \else$\mathrm{c}$\fi}
\def\bq{\ifmmode\mathrm{b}
          \else$\mathrm{b}$\fi}
\def\tq{\ifmmode\mathrm{t}
          \else$\mathrm{t}$\fi}
\def\qq{\ifmmode\mathrm{q}
          \else$\mathrm{q}$\fi}

 \def\Pgg{\ifmmode\mathrm{\gamma}
          \else$\mathrm{\gamma}$\fi}
 \def\PW{\ifmmode\mathrm{W}
         \else$\mathrm{W }$\fi}
 \def\PWp{\ifmmode\mathrm{W^+}
          \else$\mathrm{W^+}$\fi}
 \def\PWpm{\ifmmode\mathrm{W^{\pm}}
          \else$\mathrm{W^{\pm}}$\fi}
 \def\PWm{\ifmmode\mathrm{W^-}
          \else$\mathrm{W^-}$\fi}
 \def\PZz{\ifmmode\mathrm{Z^0}
          \else$\mathrm{Z^0}$\fi}
 \def\PHz{\ifmmode\mathrm{H^0}
          \else$\mathrm{H^0}$\fi}
 \def\PHpm{\ifmmode\mathrm{H^{\pm}}
           \else$\mathrm{H^{\pm}}$\fi}
 \def\PWR{\ifmmode\mathrm{W_R}
          \else$\mathrm{W_R}$\fi}
 \def\PWpr{\ifmmode\mathrm{W^{\prime}}
           \else$\mathrm{W^{\prime}}$\fi}
 \def\PZLR{\ifmmode\mathrm{Z_{LR}}
           \else$\mathrm{Z_{LR}}$\fi}
 \def\PZgc{\ifmmode\mathrm{Z_{\chi}}
           \else$\mathrm{Z_{\chi}}$\fi}
 \def\PZgy{\ifmmode\mathrm{Z_{\psi}}
           \else$\mathrm{Z_{\psi}}$\fi}
 \def\PZge{\ifmmode\mathrm{Z_{\eta}}
           \else$\mathrm{Z_{\eta}}$\fi}
 \def\PZi{\ifmmode\mathrm{Z_1}
          \else$\mathrm{Z_1}$\fi}
 \def\PAz{\ifmmode\mathrm{A^0}
          \else$\mathrm{A^0}$\fi}
 \def\Pgne{\ifmmode\mathrm{\nu_{e}}
           \else$\mathrm{\nu_{e}}$\fi}
 \def\Pagne{\ifmmode\mathrm{\overline{\nu_{e}}}
            \else$\mathrm{\overline{\nu_{e}}}$\fi}
 \def\Pgngm{\ifmmode\mathrm{\nu_{\mu}}
            \else$\mathrm{\nu_{\mu}}$\fi}
 \def\Pagngm{\ifmmode\mathrm{\overline{\nu}_{\mu}}
             \else$\mathrm{\overline{\nu}_{\mu}}$\fi}
 \def\Pgngt{\ifmmode\mathrm{\nu_{\tau}}
            \else$\mathrm{\nu_{\tau}}$\fi}
 \def\Pagngt{\ifmmode\mathrm{\overline{\nu}_{\tau}}
             \else$\mathrm{\overline{\nu}_{\tau}}$\fi}
 \def\Pe{\ifmmode\mathrm{e}
         \else$\mathrm{e}$\fi}
 \def\Pep{\ifmmode\mathrm{e^+}
          \else$\mathrm{e^+}$\fi}
 \def\Pem{\ifmmode\mathrm{e^-}
          \else$\mathrm{e^-}$\fi}
 \def\Pgm{\ifmmode\mathrm{\mu}
          \else$\mathrm{\mu}$\fi}
 \def\Pgmm{\ifmmode\mathrm{\mu^-}
           \else$\mathrm{\mu^-}$\fi}
 \def\Pgmp{\ifmmode\mathrm{\mu^+}
           \else$\mathrm{\mu^+}$\fi}
 \def\Pgt{\ifmmode\mathrm{\tau}
          \else$\mathrm{\tau}$\fi}
 \def\PLpm{\ifmmode\mathrm{L^{\pm}}
           \else$\mathrm{L^{\pm}}$\fi}
 \def\PLz{\ifmmode\mathrm{L^0}
          \else$\mathrm{L^0}$\fi}
 \def\PEz{\ifmmode\mathrm{E^0}
          \else$\mathrm{E^0}$\fi}
 \def\Pgp{\ifmmode\mathrm{\pi}
          \else$\mathrm{\pi }$\fi}
 \def\Pgpm{\ifmmode\mathrm{\pi^-}
           \else$\mathrm{\pi^-}$\fi}
 \def\Pgpp{\ifmmode\mathrm{\pi^+}
           \else$\mathrm{\pi^+}$\fi}
 \def\Pgppm{\ifmmode\mathrm{\pi^{\pm }}
            \else$\mathrm{\pi^{\pm }}$\fi}
 \def\Pgpz{\ifmmode\mathrm{\pi^0}
           \else$\mathrm{\pi^0 }$\fi}
 \def\Pgh{\ifmmode\mathrm{\eta}
          \else$\mathrm{\eta }$\fi}
 \def\Pgr{\ifmmode\mathrm{\rho(770)}
          \else$\mathrm{\rho(770)}$\fi}
 \def\Pgo{\ifmmode\mathrm{\omega(783)}
          \else$\mathrm{\omega(783)}$\fi}
 \def\Pghpr{\ifmmode\mathrm{\eta^{\prime}(958)}
            \else$\mathrm{\eta^{\prime}(958)}$\fi}
 \def\Pfz{\ifmmode\mathrm{f_0(980)}
          \else$\mathrm{f_0(980)}$\fi}
 \def\Paz{\ifmmode\mathrm{a_0(980)}
          \else$\mathrm{a_0(980)}$\fi}
 \def\Pgf{\ifmmode\mathrm{\phi(1020)}
          \else$\mathrm{\phi(1020)}$\fi}
 \def\Phia{\ifmmode\mathrm{h_1(1170)}
           \else$\mathrm{h_1(1170)}$\fi}
 \def\Pbi{\ifmmode\mathrm{b_1(1235)}
          \else$\mathrm{b_1(1235)}$\fi}
 \def\Pai{\ifmmode\mathrm{a_1(1260)}
          \else$\mathrm{a_1(1260)}$\fi}
 \def\Pfii{\ifmmode\mathrm{f_2(1270)}
           \else$\mathrm{f_2(1270)}$\fi}
 \def\Pfi{\ifmmode\mathrm{f_1(1285)}
          \else$\mathrm{f_1(1285)}$\fi}
 \def\Pgha{\ifmmode\mathrm{\eta(1295)}
           \else$\mathrm{\eta(1295)}$\fi}
 \def\Pgpa{\ifmmode\mathrm{\pi(1300)}
           \else$\mathrm{\pi(1300)}$\fi}
 \def\Paii{\ifmmode\mathrm{a_2(1320)}
           \else$\mathrm{a_2(1320)}$\fi}
 \def\Pgoa{\ifmmode\mathrm{\omega(1390)}
           \else$\mathrm{\omega(1390)}$\fi}
 \def\Pfza{\ifmmode\mathrm{f_0(1400)}
           \else$\mathrm{f_0(1400)}$\fi}
 \def\Pfia{\ifmmode\mathrm{f_1 (1390)}
           \else$\mathrm{f_1 (1390)}$\fi}
 \def\Pghb{\ifmmode\mathrm{\eta(1440)}
           \else$\mathrm{\eta(1440)}$\fi}
 \def\Pgra{\ifmmode\mathrm{\rho(1450)}
           \else$\mathrm{\rho(1450)}$\fi}
 \def\Pfib{\ifmmode\mathrm{f_1(1510)}
           \else$\mathrm{f_1(1510)}$\fi}
 \def\Pfiipr{\ifmmode\mathrm{f^{\prime}_2(1525)}
             \else$\mathrm{f^{\prime}_2(1525)}$\fi}
 \def\Pfzb{\ifmmode\mathrm{f_0(1590)}
           \else$\mathrm{f_0(1590)}$\fi}
 \def\Pgob{\ifmmode\mathrm{\omega(1600)}
           \else$\mathrm{\omega(1600)}$\fi}
 \def\Pgoiii{\ifmmode\mathrm{\omega_3(1670)}
             \else$\mathrm{\omega_3(1670)}$\fi}
 \def\Pgpii{\ifmmode\mathrm{\pi_2(1670)}
            \else$\mathrm{\pi_2(1670)}$\fi}
 \def\Pgfa{\ifmmode\mathrm{\phi(1680)}
           \else$\mathrm{\phi(1680)}$\fi}
 \def\Pgriii{\ifmmode\mathrm{\rho_3(1690)}
             \else$\mathrm{\rho_3(1690)}$\fi}
 \def\Pgrb{\ifmmode\mathrm{\rho(1700)}
           \else$\mathrm{\rho(1700)}$\fi}
 \def\Pfiia{\ifmmode\mathrm{f_2(1720)}
            \else$\mathrm{f_2(1720)}$\fi}
 \def\Pgfiii{\ifmmode\mathrm{\phi_3(1850)}
             \else$\mathrm{\phi_3(1850)}$\fi}
 \def\Pfiib{\ifmmode\mathrm{f_2(2010)}
            \else$\mathrm{f_2(2010)}$\fi}
 \def\Pfiv{\ifmmode\mathrm{f_4(2050)}
           \else$\mathrm{f_4(2050)}$\fi}
 \def\Pfiic{\ifmmode\mathrm{f_2(2300)}
            \else$\mathrm{f_2(2300)}$\fi}
 \def\Pfiid{\ifmmode\mathrm{f_2(2340)}
            \else$\mathrm{f_2(2340)}$\fi}
 \def\PK{\ifmmode\mathrm{K}
         \else$\mathrm{K}$\fi}
 \def\PKpm{\ifmmode\mathrm{K^{\pm}}
           \else$\mathrm{K^{\pm}}$\fi}
 \def\PKp{\ifmmode\mathrm{K^+}
          \else$\mathrm{K^+}$\fi}
 \def\PKm{\ifmmode\mathrm{K^-}
          \else$\mathrm{K^-}$\fi}
 \def\PKz{\ifmmode\mathrm{K^0}
          \else$\mathrm{K^0}$\fi}
 \def\PaKz{\ifmmode\mathrm{\overline{K^0}}
           \else$\mathrm{\overline{K^0}}$\fi}
 \def\PKgmiii{\ifmmode\mathrm{K_{\mu 3}}
              \else$\mathrm{K_{\mu 3}}$\fi}
 \def\PKeiii{\ifmmode\mathrm{K_{\rm e3}}
             \else$\mathrm{K_{\rm e3}}$\fi}
 \def\PKzS{\ifmmode\mathrm{K^0_{\rm S}}
           \else$\mathrm{K^0_{\rm S}}$\fi}
 \def\PKzL{\ifmmode\mathrm{K^0_{\rm L}}
           \else$\mathrm{K^0_{\rm L}}$\fi}
 \def\PKzgmiii{\ifmmode\mathrm{K^0_{\mu 3}}
               \else$\mathrm{K^0_{\mu 3}}$\fi}
 \def\PKzeiii{\ifmmode\mathrm{K^0_{{\rm e}3}}
              \else$\mathrm{K^0_{{\rm e}3}}$\fi}
 \def\PKst{\ifmmode\mathrm{K^{\ast}(892)}
           \else$\mathrm{K^{\ast}(892)}$\fi}
 \def\PKi{\ifmmode\mathrm{K_1(1270)}
          \else$\mathrm{K_1(1270)}$\fi}
 \def\PKsta{\ifmmode\mathrm{K^{\ast}(1370)}
            \else$\mathrm{K^{\ast}(1370)}$\fi}
 \def\PKia{\ifmmode\mathrm{K_1(1400)}
           \else$\mathrm{K_1(1400)}$\fi}
 \def\PKstz{\ifmmode\mathrm{K^{\ast}_0(1430)}
            \else$\mathrm{K^{\ast}_0(1430)}$\fi}
 \def\PKstii{\ifmmode\mathrm{K^{\ast}_2(1430)}
             \else$\mathrm{K^{\ast}_2(1430)}$\fi}
 \def\PKstb{\ifmmode\mathrm{K^{\ast}(1680)}
            \else$\mathrm{K^{\ast}(1680)}$\fi}
 \def\PKii{\ifmmode\mathrm{K_2(1770)}
           \else$\mathrm{K_2(1770)}$\fi}
 \def\PKstiii{\ifmmode\mathrm{K^{\ast}_3(1780)}
              \else$\mathrm{K^{\ast}_3(1780)}$\fi}
 \def\PKstiv{\ifmmode\mathrm{K^{\ast}_4(2045)}
             \else$\mathrm{K^{\ast}_4(2045)}$\fi}
 \def\PD{\ifmmode\mathrm{D}
           \else$\mathrm{D}$\fi}
 \def\PaD{\ifmmode\mathrm{\overline{ D}}
          \else${\mathrm{\overline D}}$\fi}
 \def\PDpm{\ifmmode\mathrm{D^{\pm}}
           \else$\mathrm{D^{\pm}}$\fi}
 \def\PDm{\ifmmode\mathrm{D^-}
          \else$\mathrm{D^-}$\fi}
 \def\PDp{\ifmmode\mathrm{D^+}
          \else$\mathrm{D^+}$\fi}
 \def\PDz{\ifmmode\mathrm{D^0}
          \else$\mathrm{D^0}$\fi}
 \def\PaDz{\ifmmode\mathrm{\overline{D^0}}
           \else$\mathrm{\overline{D^0}}$\fi}
 \def\PDstpm{\ifmmode{\mathrm{D}^{\ast}(2010)^{\pm}}
             \else$\mathrm{D}^{\ast}(2010)^{\pm}$\fi}
 \def\PDstp{\ifmmode{\mathrm{D}^{\ast+}}
             \else$\mathrm{D}^{\ast+}$\fi}
 \def\PDst{\ifmmode{\mathrm{D}^{\ast}}
             \else$\mathrm{D}^{\ast}$\fi}
 \def\PDstz{\ifmmode{\mathrm{D}^{\ast}(2010)^0}
            \else$\mathrm{D}^{\ast}(2010)^0$\fi}
 \def\PDiz{\ifmmode{\mathrm{D}_{1}(2420)^0}
           \else$\mathrm{D}_{1}(2420)^0$\fi}
 \def\PDstiiz{\ifmmode{\mathrm{D}^{\ast}_{2}(2460)^0}
              \else$\mathrm{D}^{\ast}_{2}(2460)^0$\fi}
 \def\PsDp{\ifmmode\mathrm{D_{s}^+}
           \else$\mathrm{D_{s}^+}$\fi}
 \def\PsDm{\ifmmode\mathrm{D_{s}^-}
           \else$\mathrm{D_{s}^-}$\fi}
 \def\PsDpm{\ifmmode\mathrm{D_{s}^{\pm}}
           \else$\mathrm{D_{s}^{\pm}}$\fi}
 \def\PsDst{\ifmmode\mathrm{D_{s}^{\ast}}
            \else$\mathrm{D_{s}^{\ast}}$\fi}
 \def\PsDipm{\ifmmode\mathrm{D_{s1}(2536)^{\pm}}
           \else$\mathrm{D_{s1}(2536)^{\pm}}$\fi}
 \def\PB{\ifmmode{\mathrm{B}}
          \else$\mathrm{B}$\fi}
 \def\PBp{\ifmmode{\mathrm{B}^{+}}
           \else$\mathrm{B}^{+}$\fi}
 \def\PBm{\ifmmode{\mathrm{B}^{-}}
           \else$\mathrm{B}^{-}$\fi}
 \def\PBpm{\ifmmode{\mathrm{B}^{\pm}}
            \else$\mathrm{B}^{\pm}$\fi}
 \def\PBz{\ifmmode{\mathrm{B}^0}
           \else$\mathrm{B}^0$\fi}
 \def\PbgL{\ifmmode{\mathrm{\Lambda}_b}
           \else$\mathrm{\Lambda}_b$\fi}
 \def\Pcgh{\ifmmode\mathrm{{\eta}_{c}(1S)}
           \else$\mathrm{{\eta}_{c}(1S)}$\fi}
 \def\PJgyy{\ifmmode\mathrm{J /\psi}
           \else$\mathrm{J /\psi}$\fi}
 \def\PJgy{\ifmmode\mathrm{J /\psi(1S)}
           \else$\mathrm{J /\psi(1S)}$\fi}
 \def\Pcgcz{\ifmmode\mathrm{{\chi}_{c0}(1P)}
            \else$\mathrm{{\chi}_{c0}(1P)}$\fi}
 \def\Pcgci{\ifmmode\mathrm{{\chi}_{c1}(1P)}
            \else$\mathrm{{\chi}_{c1}(1P)}$\fi}
 \def\Pcgcii{\ifmmode\mathrm{{\chi}_{c2}(1P)}
             \else$\mathrm{{\chi}_{c2}(1P)}$\fi}
 \def\Pgy{\ifmmode\mathrm{\psi(2S)}
          \else$\mathrm{\psi(2S)}$\fi}
 \def\Pgya{\ifmmode\mathrm{\psi(3770)}
           \else$\mathrm{\psi(3770)}$\fi}
 \def\Pgyb{\ifmmode\mathrm{\psi(4040)}
           \else$\mathrm{\psi(4040)}$\fi}
 \def\Pgyc{\ifmmode\mathrm{\psi(4160)}
           \else$\mathrm{\psi(4160)}$\fi}
 \def\Pgyd{\ifmmode\mathrm{\psi(4415)}
           \else$\mathrm{\psi(4415)}$\fi}
 \def\PgU{\ifmmode\mathrm{\Upsilon(1S)}
          \else$\mathrm{\Upsilon(1S)}$\fi}
 \def\Pbgcz{\ifmmode\mathrm{{\chi}_{b0}(1P)}
            \else$\mathrm{{\chi}_{b0}(1P)}$\fi}
 \def\Pbgci{\ifmmode\mathrm{{\chi}_{b1}(1P)}
            \else$\mathrm{{\chi}_{b1}(1P)}$\fi}
 \def\Pbgcii{\ifmmode\mathrm{{\chi}_{b2}(1P)}
             \else$\mathrm{{\chi}_{b2}(1P)}$\fi}
 \def\PgUa{\ifmmode\mathrm{\Upsilon(2S)}
           \else$\mathrm{\Upsilon(2S)}$\fi}
 \def\Pbgcza{\ifmmode\mathrm{{\chi}_{b0}(2P)}
             \else$\mathrm{{\chi}_{b0}(2P)}$\fi}
 \def\Pbgcia{\ifmmode\mathrm{{\chi}_{b1}(2P)}
             \else$\mathrm{{\chi}_{b1}(2P)}$\fi}
 \def\Pbgciia{\ifmmode\mathrm{{\chi}_{b2}(2P)}
              \else$\mathrm{{\chi}_{b2}(2P)}$\fi}
 \def\PgUb{\ifmmode\mathrm{\Upsilon(3S)}
           \else$\mathrm{\Upsilon(3S)}$\fi}
 \def\PgUc{\ifmmode\mathrm{\Upsilon(4S)}
           \else$\mathrm{\Upsilon(4S)}$\fi}
 \def\PgUd{\ifmmode\mathrm{\Upsilon(10860)}
           \else$\mathrm{\Upsilon(10860)}$\fi}
 \def\PgUe{\ifmmode\mathrm{\Upsilon(11020)}
           \else$\mathrm{\Upsilon(11020)}$\fi}
 \def\Pp{\ifmmode\mathrm{p}
         \else$\mathrm{p}$\fi}
 \def\Pap{\ifmmode\mathrm{\overline{p}}
         \else$\mathrm{\overline{p}}$\fi}
 \def\Pn{\ifmmode\mathrm{n}
         \else$\mathrm{n}$\fi}
 \def\PNa{\ifmmode\mathrm{N(1440)P_{11}}
          \else$\mathrm{N(1440)P_{11}}$\fi}
 \def\PNb{\ifmmode\mathrm{N(1520)D_{13}}
          \else$\mathrm{N(1520)D_{13}}$\fi}
 \def\PNc{\ifmmode\mathrm{N(1535)S_{11}}
          \else$\mathrm{N(1535)S_{11}}$\fi}
 \def\PNd{\ifmmode\mathrm{N(1650)S_{11}}
          \else$\mathrm{N(1650)S_{11}}$\fi}
 \def\PNe{\ifmmode\mathrm{N(1675)D_{15}}
          \else$\mathrm{N(1675)D_{15}}$\fi}
 \def\PNf{\ifmmode\mathrm{N(1680)F_{15}}
          \else$\mathrm{N(1680)F_{15}}$\fi}
 \def\PNg{\ifmmode\mathrm{N(1700)D_{13}}
          \else$\mathrm{N(1700)D_{13}}$\fi}
 \def\PNh{\ifmmode\mathrm{N(1710)P_{11}}
          \else$\mathrm{N(1710)P_{11}}$\fi}
 \def\PNi{\ifmmode\mathrm{N(1720)P_{13}}
          \else$\mathrm{N(1720)P_{13}}$\fi}
 \def\PNj{\ifmmode\mathrm{N(2190)G_{17}}
          \else$\mathrm{N(2190)G_{17}}$\fi}
 \def\PNk{\ifmmode\mathrm{N(2220)H_{19}}
          \else$\mathrm{N(2220)H_{19}}$\fi}
 \def\PNl{\ifmmode\mathrm{N(2250)G_{19}}
          \else$\mathrm{N(2250)G_{19}}$\fi}
 \def\PNm{\ifmmode\mathrm{N(2600)I_{1,11}}
          \else$\mathrm{N(2600)I_{1,11}}$\fi}
 \def\PgDa{\ifmmode\mathrm{\Delta(1232)P_{33}}
           \else$\mathrm{\Delta(1232)P_{33}}$\fi}
 \def\PgDb{\ifmmode\mathrm{\Delta(1620)S_{31}}
           \else$\mathrm{\Delta(1620)S_{31}}$\fi}
 \def\PgDc{\ifmmode\mathrm{\Delta(1700)D_{33}}
           \else$\mathrm{\Delta(1700)D_{33}}$\fi}
 \def\PgDd{\ifmmode\mathrm{\Delta(1900)S_{31}}
           \else$\mathrm{\Delta(1900)S_{31}}$\fi}
 \def\PgDe{\ifmmode\mathrm{\Delta(1905)F_{35}}
           \else$\mathrm{\Delta(1905)F_{35}}$\fi}
 \def\PgDf{\ifmmode\mathrm{\Delta(1910)P_{31}}
           \else$\mathrm{\Delta(1910)P_{31}}$\fi}
 \def\PgDh{\ifmmode\mathrm{\Delta(1920)P_{33}}
           \else$\mathrm{\Delta(1920)P_{33}}$\fi}
 \def\PgDi{\ifmmode\mathrm{\Delta(1930)D_{35}}
           \else$\mathrm{\Delta(1930)D_{35}}$\fi}
 \def\PgDj{\ifmmode\mathrm{\Delta(1950)F_{37}}
           \else$\mathrm{\Delta(1950)F_{37}}$\fi}
 \def\PgDk{\ifmmode\mathrm{\Delta(2420)H_{3,11}}
           \else$\mathrm{\Delta(2420)H_{3,11}}$\fi}
 \def\PgDpp{\ifmmode\mathrm{\Delta^{++}}
           \else$\mathrm{\Delta^{++}}$\fi}
 \def\PgL{\ifmmode\mathrm{\Lambda}
          \else$\mathrm{\Lambda}$\fi}
 \def\PagL{\ifmmode\mathrm{\overline{\Lambda}}
            \else$\mathrm{\overline{\Lambda}}$\fi}
 \def\PgLa{\ifmmode\mathrm{\Lambda(1405) S_{01}}
           \else$\mathrm{\Lambda(1405) S_{01}}$\fi}
 \def\PgLb{\ifmmode\mathrm{\Lambda(1520) D_{03}}
           \else$\mathrm{\Lambda(1520) D_{03}}$\fi}
 \def\PgLc{\ifmmode\mathrm{\Lambda(1600) P_{01}}
           \else$\mathrm{\Lambda(1600) P_{01}}$\fi}
 \def\PgLd{\ifmmode\mathrm{\Lambda(1670) S_{01}}
           \else$\mathrm{\Lambda(1670) S_{01}}$\fi}
 \def\PgLe{\ifmmode\mathrm{\Lambda(1690) D_{03}}
           \else$\mathrm{\Lambda(1690) D_{03}}$\fi}
 \def\PgLf{\ifmmode\mathrm{\Lambda(1800) S_{01}}
           \else$\mathrm{\Lambda(1800) S_{01}}$\fi}
 \def\PgLg{\ifmmode\mathrm{\Lambda(1810) P_{01}}
           \else$\mathrm{\Lambda(1810) P_{01}}$\fi}
 \def\PgLh{\ifmmode\mathrm{\Lambda(1820) F_{05}}
           \else$\mathrm{\Lambda(1820) F_{05}}$\fi}
 \def\PgLi{\ifmmode\mathrm{\Lambda(1830) D_{05}}
           \else$\mathrm{\Lambda(1830) D_{05}}$\fi}
 \def\PgLj{\ifmmode\mathrm{\Lambda(1890) P_{03}}
           \else$\mathrm{\Lambda(1890) P_{03}}$\fi}
 \def\PgLk{\ifmmode\mathrm{\Lambda(2100) G_{07}}
           \else$\mathrm{\Lambda(2100) G_{07}}$\fi}
 \def\PgLl{\ifmmode\mathrm{\Lambda(2110) F_{05}}
           \else$\mathrm{\Lambda(2110) F_{05}}$\fi}
 \def\PgLm{\ifmmode\mathrm{\Lambda(2350) H_{09}}
           \else$\mathrm{\Lambda(2350) H_{09}}$\fi}
 \def\PgS{\ifmmode{\rm \Sigma}
           \else${\rm \Sigma}$\fi}
 \def\PgSp{\ifmmode\mathrm{\Sigma^+}
           \else$\mathrm{\Sigma^+}$\fi}
 \def\PgSz{\ifmmode\mathrm{\Sigma^0}
           \else$\mathrm{\Sigma^0}$\fi}
 \def\PgSm{\ifmmode\mathrm{\Sigma^-}
           \else$\mathrm{\Sigma^-}$\fi}
 \def\PgSpm{\ifmmode\mathrm{\Sigma^{\pm}}
           \else$\mathrm{\Sigma^{\pm}}$\fi}
 \def\PgSa{\ifmmode\mathrm{\Sigma(1385) P_{13}}
           \else$\mathrm{\Sigma(1385) P_{13}}$\fi}
 \def\PgSb{\ifmmode\mathrm{\Sigma(1660) P_{11}}
           \else$\mathrm{\Sigma(1660) P_{11}}$\fi}
 \def\PgSc{\ifmmode\mathrm{\Sigma(1670) D_{13}}
           \else$\mathrm{\Sigma(1670) D_{13}}$\fi}
 \def\PgSd{\ifmmode\mathrm{\Sigma(1750) S_{11}}
           \else$\mathrm{\Sigma(1750) S_{11}}$\fi}
 \def\PgSe{\ifmmode\mathrm{\Sigma(1775) D_{15}}
           \else$\mathrm{\Sigma(1775) D_{15}}$\fi}
 \def\PgSf{\ifmmode\mathrm{\Sigma(1915) F_{15}}
           \else$\mathrm{\Sigma(1915) F_{15}}$\fi}
 \def\PgSg{\ifmmode\mathrm{\Sigma(1940) D_{13}}
           \else$\mathrm{\Sigma(1940) D_{13}}$\fi}
 \def\PgSh{\ifmmode\mathrm{\Sigma(2030) F_{17}}
           \else$\mathrm{\Sigma(2030) F_{17}}$\fi}
 \def\PgSi{\ifmmode\mathrm{\Sigma(2050)}
           \else$\mathrm{\Sigma(2050)}$\fi}

\def\PgX{\ifmmode\mathrm{\Xi}
          \else$\mathrm{\Xi}$\fi}
\def\PagX{\ifmmode\mathrm{\overline{\Xi}}
            \else$\mathrm{\overline{\Xi}}$\fi}
 \def\PgXz{\ifmmode\mathrm{\Xi^0}
           \else$\mathrm{\Xi^0}$\fi}
 \def\PgXm{\ifmmode\mathrm{\Xi^-}
           \else$\mathrm{\Xi^-}$\fi}
 \def\PgXa{\ifmmode\mathrm{\Xi(1530)}
           \else$\mathrm{\Xi(1530)}$\fi}
 \def\PgXas{\ifmmode\mathrm{\Xi(1530)P_{13}}
           \else$\mathrm{\Xi(1530)P_{13}}$\fi}
 \def\PgXb{\ifmmode\mathrm{\Xi(1690)}
           \else$\mathrm{\Xi(1690)}$\fi}
 \def\PgXbb{\ifmmode\mathrm{\Xi(1620)}
           \else$\mathrm{\Xi(1620)}$\fi}
 \def\PgXc{\ifmmode\mathrm{\Xi(1820)D_{13}}
           \else$\mathrm{\Xi(1820)D_{13}}$\fi}
 \def\PgXcs{\ifmmode\mathrm{\Xi(1820)}
           \else$\mathrm{\Xi(1820)}$\fi}
 \def\PgXd{\ifmmode\mathrm{\Xi(1950)}
           \else$\mathrm{\Xi(1950)}$\fi}
 \def\PgXe{\ifmmode\mathrm{\Xi(2030)}
           \else$\mathrm{\Xi(2030)}$\fi}
 \def\PgOm{\ifmmode\mathrm{\Omega^-}
           \else$\mathrm{\Omega^-}$\fi}
 \def\PgO{\ifmmode\mathrm{\Omega}
           \else$\mathrm{\Omega}$\fi}
 \def\PgOma{\ifmmode\mathrm{\Omega(2250)^-}
            \else$\mathrm{\Omega(2250)^-}$\fi}
 \def\PcgL{\ifmmode\mathrm{\Lambda_c}
            \else$\mathrm{\Lambda_c}$\fi}
 \def\PacgL{\ifmmode\mathrm{\overline{\Lambda}_c}
            \else$\mathrm{\overline{\Lambda}_c}$\fi}
 \def\PcgLp{\ifmmode\mathrm{\Lambda_c^+}
            \else$\mathrm{\Lambda_c^+}$\fi}
 \def\PcgLm{\ifmmode{\rm \Lambda_c^-}
            \else${\rm \Lambda_c^-}$\fi}
 \def\PcgX{\ifmmode\mathrm{\Xi_c}
            \else$\mathrm{\Xi_c}$\fi}
 \def\PcgXz{\ifmmode\mathrm{\Xi_c^0}
            \else$\mathrm{\Xi_c^0}$\fi}
 \def\PcgXp{\ifmmode\mathrm{\Xi_c^+}
            \else$\mathrm{\Xi_c^+}$\fi}
 \def\PcgS{\ifmmode\mathrm{\Sigma_c}
           \else$\mathrm{\Sigma_c}$\fi}
 \def\PcgSz{\ifmmode\mathrm{\Sigma_c^0}
           \else$\mathrm{\Sigma_c^0}$\fi}
 \def\PcgSp{\ifmmode\mathrm{\Sigma_c^+}
           \else$\mathrm{\Sigma_c^+}$\fi}
 \def\PcgSpp{\ifmmode\mathrm{\Sigma_c^{++}}
           \else$\mathrm{\Sigma_c^{++}}$\fi}
 \def\PcgO{\ifmmode{\mathrm \Omega_c}
           \else${\mathrm \Omega_c}$\fi}
 \def\PcgOz{\ifmmode{\mathrm \Omega_c^{0}}
           \else${\mathrm \Omega_c^{0}}$\fi}
 \def\PSgg{\ifmmode\mathrm{\tilde{\gamma}}
           \else$\mathrm{\tilde{\gamma}}$\fi}
 \def\PSgxz{\ifmmode\mathrm{\tilde{\chi}^0_i}
            \else$\mathrm{\tilde{\chi}^0_i}$\fi}
 \def\PSZz{\ifmmode\mathrm{\tilde{Z}^0}
           \else$\mathrm{\tilde{Z}^0}$\fi}
 \def\PSHz{\ifmmode\mathrm{\tilde{H}^0_j}
           \else$\mathrm{\tilde{H}^0_j}$\fi}
 \def\PSgxpm{\ifmmode\mathrm{\tilde{\chi}^{\pm_i}}
             \else$\mathrm{\tilde{\chi}^{\pm_i}}$\fi}
 \def\PSWpm{\ifmmode\mathrm{\tilde{W}^{\pm}}
            \else$\mathrm{\tilde{W}^{\pm}}$\fi}
 \def\PSHpm{\ifmmode\mathrm{\tilde{H}^{\pm_j}}
            \else$\mathrm{\tilde{H}^{\pm_j}}$\fi}
 \def\PSgn{\ifmmode\mathrm{\tilde{\nu}}
           \else$\mathrm{\tilde{\nu}}$\fi}
 \def\PSe{\ifmmode\mathrm{\tilde{e}}
          \else$\mathrm{\tilde{e}}$\fi}
 \def\PSgm{\ifmmode\mathrm{\tilde{\mu}}
           \else$\mathrm{\tilde{\mu}}$\fi}
 \def\PSgt{\ifmmode\mathrm{\tilde{\tau}}
           \else$\mathrm{\tilde{\tau}}$\fi}
 \def\PSq{\ifmmode\mathrm{\tilde{q}}
          \else$\mathrm{\tilde{q}}$\fi}
 \def\PSg{\ifmmode\mathrm{\tilde{g}}
          \else$\mathrm{\tilde{g}}$\fi}

\def\tp{\ifmmode\mathrm{\Theta^+}
          \else$\mathrm{\Theta^+}$\fi}
\def\t1540{\ifmmode{\mathrm \Theta(1540)^+}
           \else${\mathrm \Theta(1540)^+}$\fi}
\def\p1860{\ifmmode{\mathrm \Phi(1860)^{--}}
           \else${\mathrm \Phi(1860)^{--}}$\fi}

\def\pbar{\ifmmode\mathrm{\overline{p}}
         \else$\mathrm{\overline{p}}$\fi}
\def\pbarp{\ifmmode\mathrm{\overline{p}p}
         \else$\mathrm{\overline{p}p}$\fi}
\def\pbarn{\ifmmode\mathrm{\overline{p}n}
         \else$\mathrm{\overline{p}n}$\fi}

\def\Cascade{\ifmmode\mathrm{\Xi}
           \else$\mathrm{\Xi}$\fi}
\def\Cascadebar{\ifmmode\mathrm{\overline{\Xi}}
           \else$\mathrm{\overline{\Xi}}$\fi}

\def\Kp{\ifmmode\mathrm{K^+}
          \else$\mathrm{K^+}$\fi}
\def\Km{\ifmmode\mathrm{K^-}
          \else$\mathrm{K^-}$\fi}

\def\Panda{{\sc{Panda}}}

\title{Studies of Hyperons and Antihyperons in Nuclei}

\ShortTitle{Studies of Hyperons and Antihyperons in Nuclei}

\author{\speaker{Josef POCHODZALLA}\\
        Johannes Gutenberg-Universit\"at Mainz, Institut f\"ur Kernphysik, D-55099 Germany\\
        E-mail: \email{pochodza@kph.uni-mainz.de}}
\author{Alexander BOTVINA\\
        Institute for Nuclear Research, Russian Academy of Sciences, 117312 Moscow, Russia\\
        E-mail: \email{a.botvina@gsi.de}}
\author{Alicia SANCHEZ LORENTE\\
        Johannes Gutenberg-Universit\"at Mainz, Institut f\"ur Kernphysik, D-55099 Germany\\
        E-mail: \email{lorente@kph.uni-mainz.de}}

\abstract{
Stored antiproton beams at the international FAIR facility will provide unique opportunities
to study hyperons as well as antihyperons in nuclear systems.
Precise $\gamma$-spectroscopy of multi-strange hypernuclei will serve as a
laboratory for the hyperon-hyperon interaction.
Exclusive hadron-antihadron pair production close to threshold
can measure the potential of a antihadron relative to that of the coincident hadrons.

In the present work we explore the production of excited states in double
hypernuclei following the micro-canonical break-up of an initially excited
double hypernucleus which is created  by the absorption and conversion of a stopped
$\Xi^{-}$ hyperon. Generally the formation of
excited hypernuclear states relative to ground states dominates in this model. For different initial
target nuclei which absorb the  $\Xi^-$,  different double hypernuclei
nuclei dominate. We also compare the model predictions with
the correlated pion spectra measured by the E906 collaboration.

In antiproton nucleus reactions the event-by-event transverse momentum
correlations of hadron-antihadron pairs produced close to threshold
contain information on the difference between the nuclear potential of the
hadron and the associated antihadron. For produced D-meson pairs at 6.7\gevc1
the sensitivity of the transverse momenta correlation will probably
be to small to deduce differences between the potentials for D$^+$
and D$^-$ mesons. However, for {\PgX\PagX} pairs produced at
2.9\gevc1 the asymmetry is sufficiently sensitive to predicted
differences between the {\PgX} and {\PagX} potentials even if the momentum
and density dependence of the the potential are taken into account.
}

\FullConference{XLVIII International Winter Meeting on Nuclear Physics in Memoriam of Ileana Iori \\
25-29 January 2010\\
Bormio, Italy}

\begin{document}

\section{Introduction}
Quantum Chromo Dynamics (QCD) is the theory of the force responsible for the binding of nucleons and nuclei and thus of a significant fraction of the ordinary matter in our universe. While the internal structure of hadrons and the spectra of their excited states are important aspects of QCD, it is at least equally important to understand how nuclear physics emerges in a more rigorous way out of QCD and how nuclear structures - nuclei on the small scale and dense stellar objects on the large scale - are formed. In particular the role of strangeness in neutron stars has not been settled yet.

The study of hypernuclei can illuminate features that are obscured in conventional nuclear systems. The hyperon offers a selective probe of the hadronic many-body problem as it is not restricted by the Pauli principle. Since direct scattering experiments between two hyperons are impractical, the precise
spectroscopy of multi-strange hypernuclei provides a unique chance
to explore the hyperon-hyperon interaction. Significant progress in nuclear structure calculations in chiral effective field theory nurtures the hope that detailed information on excitation spectra of double hypernuclei will provide unique information on the hyperon-hyperon interactions. At the same time, the  $\Lambda$-N and $\Lambda$-$\Lambda$ weak interaction can be studied by hypernuclei decays, opening a inimitable window for the four-baryon, strangeness non-conserving weak interaction.

Furthermore, the physics of strangeness in hadronic systems is a constantly developing field. It brings up new, often unexpected results, new challenges and open questions like anti-hypernuclei, kaonic nuclei, exotic hadronic states like the controversially discussed pentaquark baryons and the H-dibaryon. Based on G-parity transformation~\cite{Lee56} D\"urr and Teller predicted within an early form of a relativistic field theory a strongly attractive potential for antiprotons in nuclei~\cite{Duer56a}. First investigations of
antiproton-nucleus scattering cross sections~\cite{Duer58,Gol58b,Ben94}
showed however disagreement with a strong attractive potential.
Later, X-ray transitions in antiprotonic atoms~\cite{Bar72,Bac72,Rob77,Pot78,Fri07} gave also hints for an
attractive potential albeit with large
 uncertainties~\cite{Aue81,Bat81}. More comprehensive studies
\cite{Fri05} of antiprotonic X-rays  as well as recent analyses of
the production of antiprotons in reactions with heavy ions resulted
in attractive real potentials in the range of about -100 to
-150~MeV~\cite{Teis94,Spie96,Sib98}. It is obvious that G-parity can
only provide a link between the $NN$ and $N\overline{N}$
interactions for distances where meson exchange is a valid
concept~\cite{Dov80,Fae82}. For distances lower than about 1~fm,
quark degrees of freedom may play a decisive role. Nevertheless this
consideration still indicates that a direct comparison of the
interactions of baryons and antibaryons in nuclei may help to shed
some light on the nature of short-range baryon-baryon forces.

\begin{figure}[t]
\begin{center}
\includegraphics[width=0.75\linewidth]{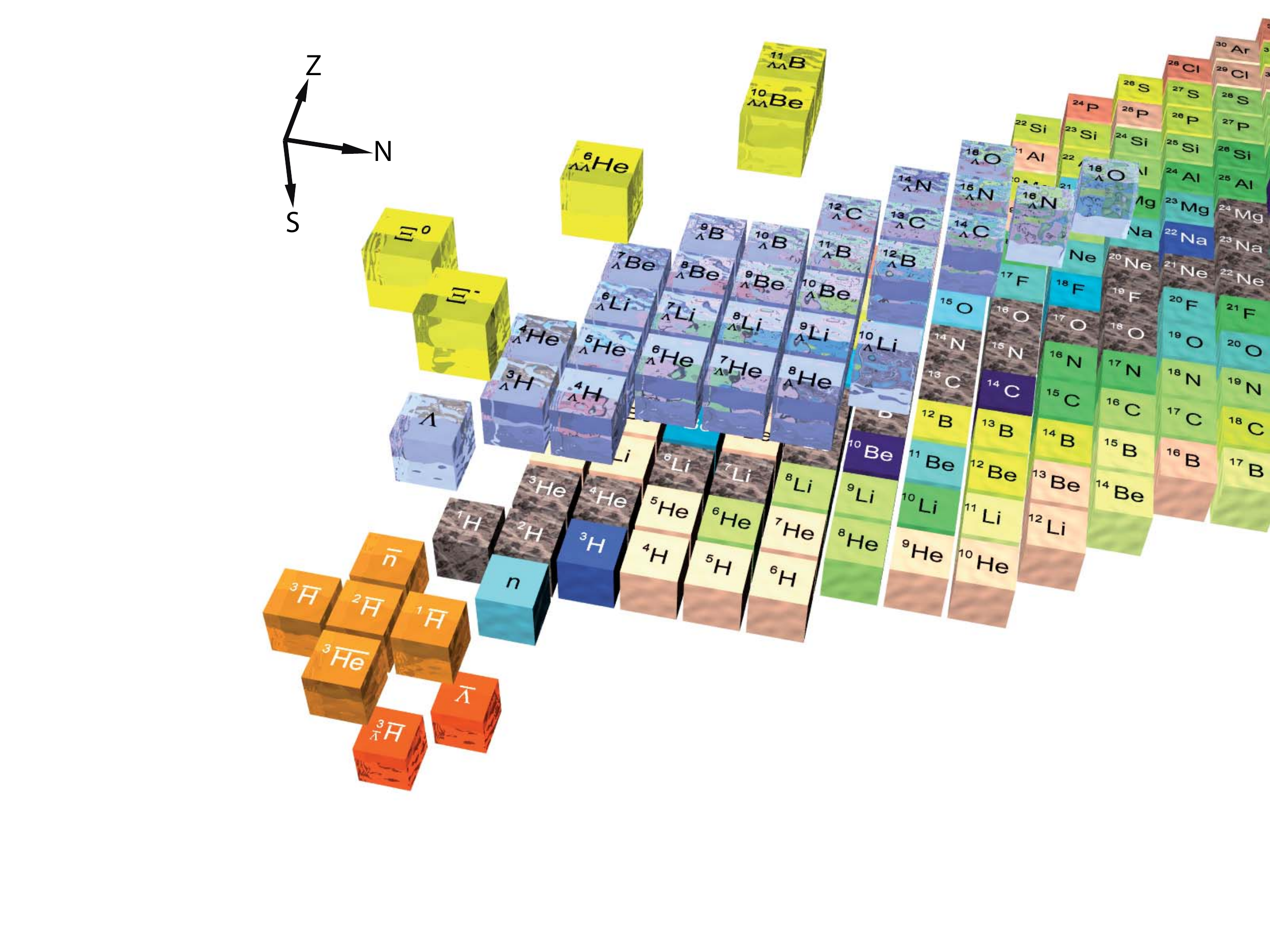}
\end{center}
\caption{Present knowledge on S=-1 nuclei (blue) and S=-2 nuclei (yellow). Only very
  few individual events of double hypernuclei have been detected and identified so far.
  First antihypertritons were recently observed by the STAR collaboration \cite{STAR}.
}
\label{fig:bormio01}
\end{figure}

\section{Production of Double Hypernuclei}
The simultaneous production and implementation of two $\Lambda$
particles into a nucleus is intricate. There is a
possibility to produce multi-strange hypernuclei in heavy ion
collisions via coalescence \cite{Ker73,Bot07}. The first observation of antihypernuclei by the
STAR collaboration impressively illustrates the potential of this method \cite{STAR}. However, high
resolution spectroscopy of excited states is not feasible. To
produce double hypernuclei in a more `controlled' way the conversion
of a captured $\Cascade^-$ and a proton into two $\Lambda$ particles
can be used. This process releases -- ignoring binding energy
effects -- only 28\,MeV. For light nuclei there exists therefore a
significant probability of the order of a few percent that both
$\Lambda$ hyperons are trapped in the same excited nuclear
fragment~\cite{Yam94,Yam97,Yam97b,Hir97,Hir99,Aok09}.

Unfortunately $\Cascade^-$ hyperons produced in reactions with
stable hadron beams have usually rather high momenta. Therefore,
direct capture of the $\Cascade^-$ in the nucleus is rather
unlikely. Even in case of the $(\Km, \Kp)$ double strangeness
exchange reaction, $\Cascade^-$ hyperons are produced with typical
momenta of 500 {\mevc1} at beam momenta around 1.8 GeV/c
\cite{Yam97b,Ohn01}. The advantage of this production process is that
the outgoing $\Kp$ can be used as a tag for the reaction. A drawback
is the low kaon beam intensity and hence the need for thick primary
targets. Furthermore, as a consequence of the large momentum
transfer, the probability to form bound $\Xi^-$ states directly is
rather small on the level of 1$\%$ \cite{Ike94,Yam94b} and the
production of quasi-free  $\Xi^-$ dominates. Still the $\Xi^-$
hyperons in the quasi-free region may be absorbed into the target
nucleus via a rescattering process on a nucleon which itself is
knocked out of the primary nucleus. This two-step process is
predicted to exceed the direct capture by more than a factor of 6
\cite{Yam97}.

On the other hand most ($\sim 80\% $) $\Xi^-$ hyperons escape from
the primary target nucleus in $(\Km, \Kp)$ reactions. However, in a
second step, these $\Cascade^-$ hyperons can be slowed down in a
dense, solid material (e.g.\ a nuclear emulsion) and form
$\Cascade^-$ atoms~\cite{Bat99}. After an atomic cascade, the
$\Xi$-hyperon is eventually captured by a secondary target nucleus
and converted via the $\Xi^-p\rightarrow \Lambda\Lambda$ reaction
into two $\Lambda$ hyperons. In a similar two-step process
relatively low momentum $\Cascade^-$ can also be produced using antiproton beams in $\pbarp
\rightarrow \Cascade^- \Cascadebar^+$ or $\pbarn \rightarrow
\Cascade^- \Cascadebar^\circ$ reactions if this reactions happens in
a complex nucleus where the produced $\Xi^-$ can re-scatter
\cite{Poc04,PandaTPR}. The advantage as compared to the kaon induced
reaction is that antiprotons are stable and can be
retained in a storage ring. This allows a rather high luminosity
even with very thin primary targets.

Because of the two-step mechanism, spectroscopic studies, based on
two-body kinematics like in single hypernucleus production, cannot
be performed. Spectroscopic information on double hypernuclei can
therefore only be obtained via their decay products. The kinetic
energies of weak decay products are sensitive to the binding
energies of the two $\Lambda$ hyperons.
While the double pionic decay of light double hypernuclei can be used
as an effective experimental filter to reduce the background
\cite{PandaPhysicsbook}
the unique identification of hypernuclei groundstates only via their pionic decay is usually hampered by the limited resolution. Instead, $\gamma$-rays emitted via the sequential decay of excited double hypernuclei may provide precise
information on the level structure.

\subsection{Statistical Decay of excited Doubly Strange Nuclei}

The \Panda\ experiment \cite{PandaTPR} which is planned at the
international Facility for Antiproton and Ion Research {\sc FAIR} in
Darmstadt aims at the high resolution $\gamma$-ray spectroscopy of
double hypernuclei \cite{Poc04}. An important question is to what extent
double hypernuclei in excited, particle stable states are populated
following the break-up of an highly excited doubly strange nucleus which is formed after the
absorption and conversion of a stopped $\Xi ^{-}$.
For light nuclei even a relatively small excitation energy may  be comparable with
their binding energy. We therefore consider the explosive decay of
the excited nucleus into several smaller clusters as the principal
mechanism of de-excitation.

To describe this break-up process we have developed a model \cite{sanchez10} which is
similar to the famous Fermi model for particle production in nuclear
reactions \cite{Fermi}. We assume that the nucleus with mass numbers
$A_0$, charge $Z_0$, and the number of $\Lambda$ hyperons $H_0$
(here $H_0$=2) breaks-up simultaneously into cold and slightly
excited fragments, which have a lifetime longer than the decay time,
estimated as an order of 100-300 fm/c. In the model we consider all possible break-up channels, which
satisfy the mass number, hyperon number (i.e. strangeness), charge,
energy and momentum conservations, and take into account the
competition between these channels.

The excitation energy of the initial, highly excited double $\Lambda$ nucleus is determined
by the binding energy of the captured $\Xi^-$ hyperon.
Unfortunately, still very little is established experimentally on
the interaction of $\Xi^-$ hyperons with nuclei. Various data suggest
a nuclear potential well depth around 20\,MeV (see e.g.
\cite{Dov83,Fri07}). Calculations of light $\Xi$ atoms \cite{Bat99}
predict that the conversion of the captured $\Xi^-$ from excited
states with correspondingly small binding energies dominates. In a
nuclear emulsion experiment a $\Xi^-$ capture at rest with two
single hyperfragments has been observed \cite{Aok95} which was
interpreted as $\Xi^-$ + $^{12}$C $\rightarrow ^4_{\Lambda}$H +
$^9_{\Lambda}$Be reaction. The deduced binding energy of the $\Xi^-$
varied between 0.62\,MeV and 3.70\,MeV, depending whether only one
out of the two hyperfragments or both fragments were produced in an
excited particle stable state. In
order to take into account the uncertainties of the excitation
energy of the converted $\Xi^-$-states, the calculations were
performed for a range of energies $0 \leq B_{\Xi} \leq E_{max}$,
constructing in this way the excitation functions for the production
of hypernuclei.

\begin{figure}[t]
\begin{center}
\includegraphics[width=0.57\linewidth]{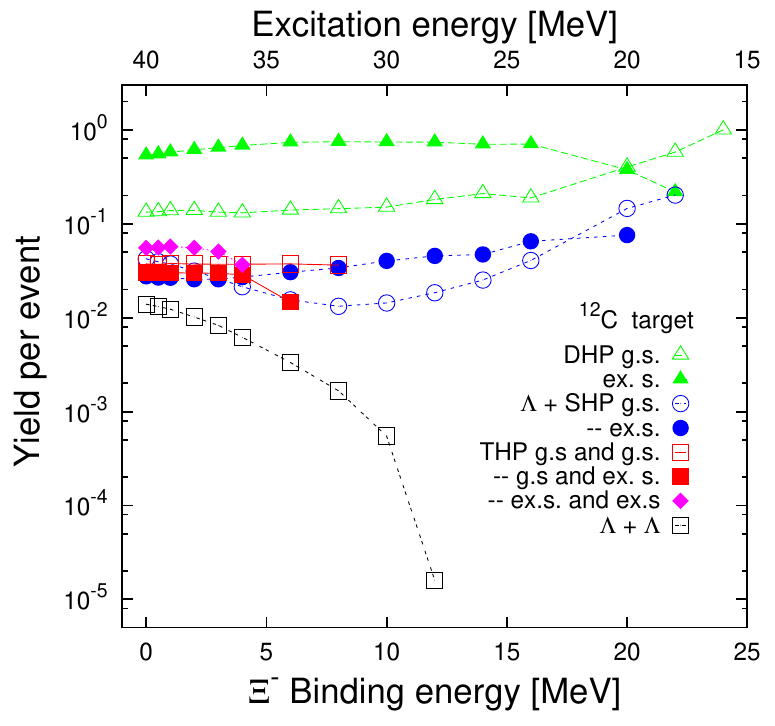}
\end{center}
\caption{
Predicted production probability of ground (g.s.) and excited states
(ex.s.) in one single (SHP), twin (THP) and double hypernuclei (DHP)
after the capture of a $\Xi^-$ in a $^{12}$C nucleus and its
conversion into two $\Lambda$ hyperons . The lower and upper scale shows the binding energy of the captured $\Xi^-$ and the excitation energy of the initial $^{13}_{\Lambda\Lambda}$B nucleus, respectively (from Ref.~\cite{sanchez10}).}
\label{fig:bormio02}
\end{figure}

Fig.~\ref{fig:bormio02} shows as an example the production of
ground (g.s.) and excited (ex.s.) states of single + one free $\Lambda$ (SHP), twin (THP)
and double (DHP) hypernuclei in case of a $^{12}$C target as a
function of the assumed $\Xi^-$ binding energy.
With increasing $\Xi^-$ binding energy the excitation energy of the excited primary $^{13}_{\Lambda\Lambda}$B$^*$ nucleus decreases from left to right from about 40\,MeV to 15\,MeV. For all excitation energies above 20\,MeV the production of {\em
excited} double hypernuclei dominates (green triangles). This can be
traced back to the opening of several thresholds for various excited
double hypernuclei already at moderate excitation energies. Only for small binding energies and
hence large excitation energies the production of single and twin
hypernuclei is significant ($\sim$10$\%$). The
non-monotonic behaviour for single hypernucleus + one free $\Lambda$
production reflects the fact that the various lowest thresholds are
relative high and widely separated, e.g. $^{12}_{\Lambda}$B+$\Lambda$
at B$_{\Xi}$=23.9\,MeV followed by $^{11}_{\Lambda}$B+n+$\Lambda$ at
11.3\,MeV. Twin-hypernuclei are only
produced for $\Xi^-$ binding energies below the threshold for
$^8_{\Lambda}$Li + $^5_{\Lambda}$He with B$_{\Xi}$=13.6\,MeV. As
discussed above the frequent observation of twin-hypernuclei
\cite{Wil59,Ste63,Bec68,Aok93,Aok95,Ich01} signals a conversion from
a $\Xi$ state with only moderate binding energy. In this range of
B$_{\Xi}$ the production probability of double hypernuclei is
comparable to previous estimates within a canonical statistical
model \cite{Yam97,Yam97b}. It is important to stress that these
numbers do not include a possible pre-equilibrium emission of
hyperons during the capture and conversion stage. With respect to the
number of stopped $\Xi^-$ hyperons, pre-equilibrium processes will
decrease the yield of double hypernuclei relative to
the yield for single hypernuclei (see e.g. \cite{Aok09}). Indeed in the
simulations for the planned \Panda\ experiment \cite{PandaPhysicsbook}
a joint capture$\times$conversion probability of 5$\%$ was assumed
to mimic this pre-equilibrium stage.

Using different $\Xi$-absorbing stable secondary target
nuclei $^9$Be, $^{10}$B, $^{11}$B, $^{12}$C and $^{13}$C one finds that
different double hypernuclei dominate for each target \cite{sanchez10}.
Thus combining the information shown in Fig.~\ref{fig:bormio02} with the measurement of
two pion momenta from the subsequent weak decays a unique assignment of various newly observed
$\gamma$-transitions to specific double hypernuclei seems possible as intended by the {\Panda} collaboration \cite{Poc04,PandaPhysicsbook}.

\subsection{The E906 Puzzle}
\begin{figure}[t]
\begin{center}
\includegraphics[width=0.7\linewidth]{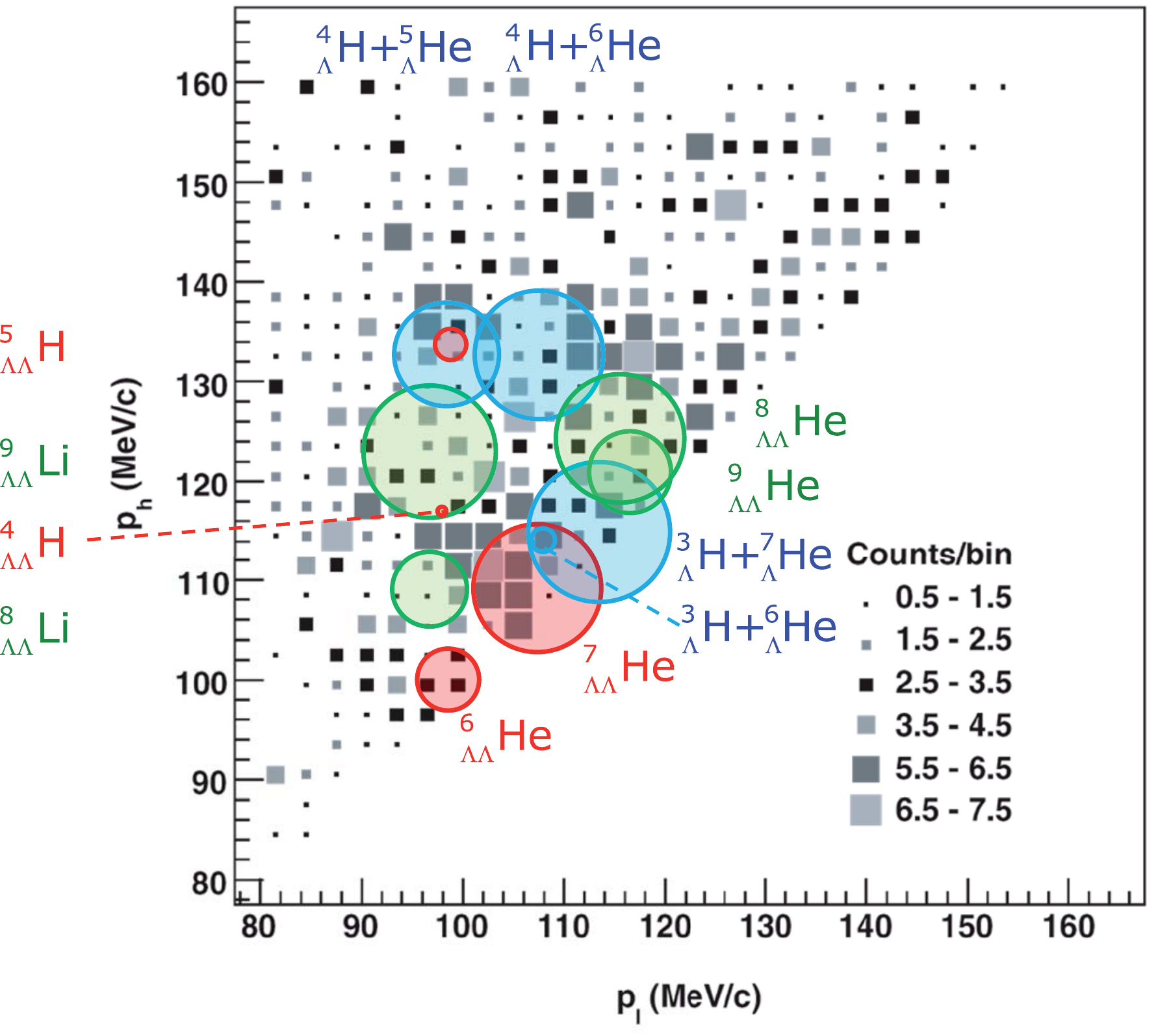}\end{center}
\caption{Momenta of two correlated pions measured by the E906 collaboration (grey squares).
Overlaid are production yields of various twin pairs (blue), double hypernuclei including their excited states (green) and double hypernuclei with a stable groundstate only (red). The areas of the circles are proportional to the {\em production} yields predicted by the statistical model. Pionic decay probabilities are not included in this plot.
}
\label{fig:bormio03}

\end{figure}

In 2001 the BNL experiment E906 reported the observation of the
$^4_{\Lambda\Lambda}$H hypernucleus by measuring the sequential
pionic decays after a (K$^-$,K$^+$) reaction deposited two units of
strangeness in a $^9$Be target \cite{Ahn01} (see Fig.~\ref{fig:bormio03}). Two structures
in the correlated $\pi^-$ momenta at (133,114)\mevc1 and at (114,104)\mevc1
were observed. The first structure was interpreted as the production of
$^3_{\Lambda}$H+$^4_{\Lambda}$H twins while the bump at (114,104)\mevc1 was attributed to pionic decays of the double hypernucleus $^4_{\Lambda\Lambda}$H. However, as it was pointed out by Kumagai-Fuse and Okabe also twin $\Lambda$-hypernuclear decays of $^3_{\Lambda}$H and $^6_{\Lambda}$He are a possible candidate to form this peak if
excited resonance states of $^6$Li are considered \cite{Kum02}. More
recently Randeniya and Hungerford showed that the published E906 data can be reproduced without the inclusion of $^4_{\Lambda\Lambda}$H decay and that it is
more likely that the decay of $^7_{\Lambda\Lambda}$He was observed
in the E906 experiment \cite{Ran07}. In their analysis this double
hypernucleus was accompanied by a background of coincident decays of single hypernuclei pairs $^3_{\Lambda}$H+$^4_{\Lambda}$H, $^3_{\Lambda}$H+$^3_{\Lambda}$H, and
$^4_{\Lambda}$H+$^4_{\Lambda}$H, respectively.

\begin{figure}[t]
\begin{center}
\includegraphics[width=0.85\linewidth]{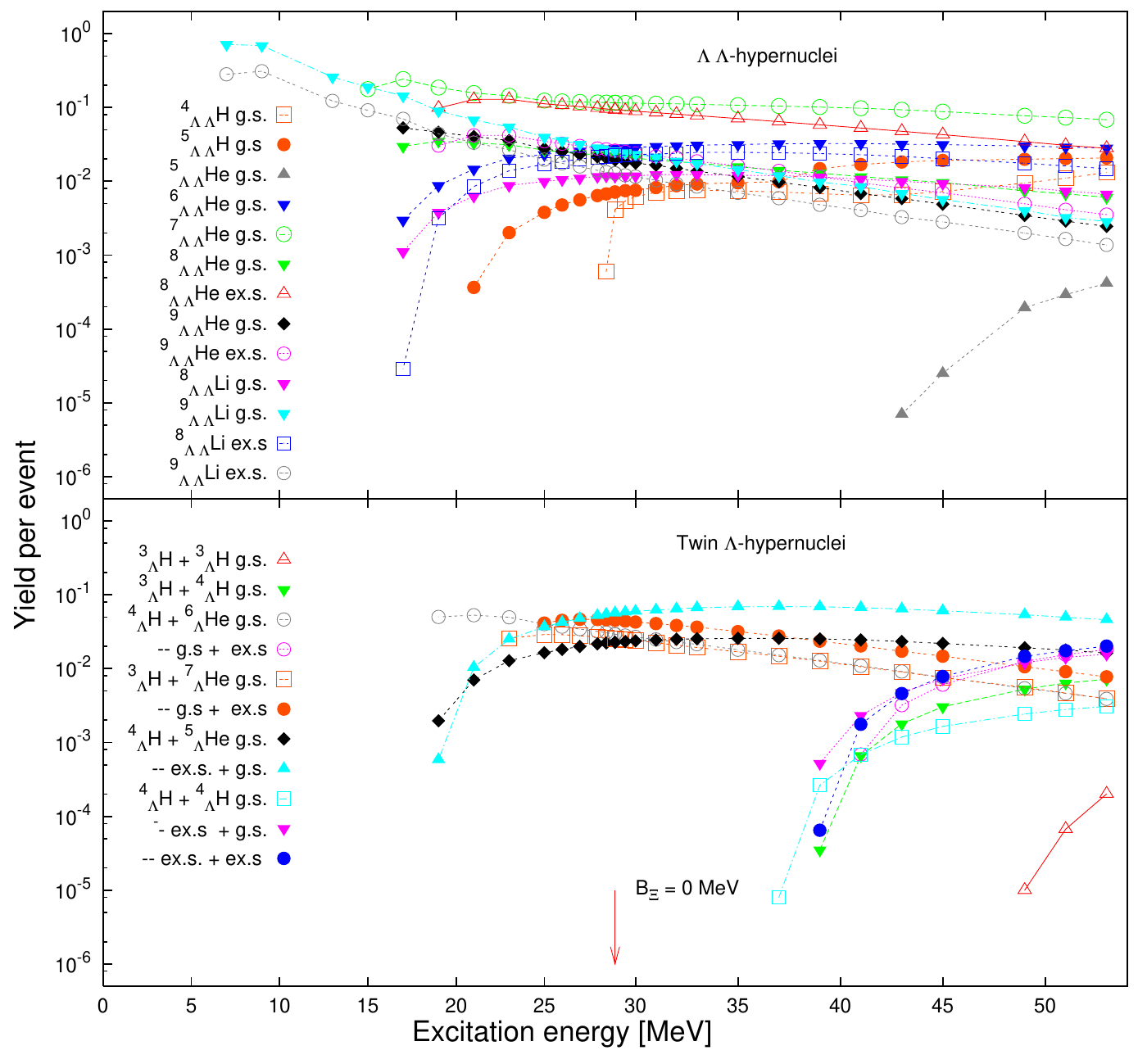}\end{center}
\caption{Relative production probability of double (top part) and twin hypernuclei (lower part) for a primary $^{10}_{\Lambda\Lambda}$Li nucleus as a function of its excitation energy.}
\label{fig:bormio04}
\end{figure}

Fig. \ref{fig:bormio04} shows the predicted relative probabilities for
the  production of particle stable twin and double hypernuclei in the
E906 experiment after the capture and conversion of a {\em stopped}
$\Xi^-$ in a secondary $^9$Be target $\Xi^-+^9$Be$\rightarrow ^{10}_{\Lambda\Lambda}$Li$^*$. As before a $\Xi^-$ binding energy of 0.5\,MeV seems reasonable corresponding to an $^{10}_{\Lambda\Lambda}$Li excitation of about 29\,MeV.
The produced yields for this excitation energy are shown in Fig.~\ref{fig:bormio03}. Here, ground state and excited state(s) - if they exist - have been added and the pion momenta for groundstate decays are assumed.
Note, that the production of $^4_{\Lambda}$H+$^4_{\Lambda}$H twins is even at an excitation energy of 35\,MeV energetically not possible and - unlike to a canonical calculations \cite{Yam97b} - does therefore not occur in our micro-canonical model.

Let us first discuss the structure at (114,104){\mevc1} which has been attributed to double hypernuclei decays. Generally the production of double hypernuclei is energetically favored over the production of twins: all possible channels with twin-production lie energetically significantly above the thresholds for
$^9_{\Lambda\Lambda}$Li, $^7_{\Lambda\Lambda}$He and
$^6_{\Lambda\Lambda}$He production in case of the $^{10}_{\Lambda\Lambda}$Li compound picture. Correspondingly the production of $^3_{\Lambda}$H and $^6_{\Lambda}$He twins which has been suggested  as a possible source of the peak structure around (114,104)\mevc1 \cite{Kum02}, is in our model by a factor of 30 lower than the $^7_{\Lambda\Lambda}$He production probability.
Unlike in the canonical model of Ref.~\cite{Yam97b} the $^7_{\Lambda\Lambda}$He probability exceeds also the one of a $^4_{\Lambda\Lambda}$H by more than two orders of magnitude and the production of $^5_{\Lambda\Lambda}$H by a factor of about 17 in our model. Of course, a direct comparison with the E906 data requires a detailed consideration of the branching ratios for pionic two-body decays many of which are not known so far.
Keeping that caveat in mind our microcanonical model supports independently on the assumed production scheme the interpretation of the E906 observation by Randeniya and Hungerford \cite{Ran07} in terms of $^7_{\Lambda\Lambda}$He decays. Decays from the ground or even exited states of $^8_{\Lambda\Lambda}$Li or $^9_{\Lambda\Lambda}$Li can possibly contribute some background to the structure at (114,104){\mevc1}.

In order to describe the (133,114)\mevc1 structure of the E906 experiment, the production of $^3_{\Lambda}$H+$^4_{\Lambda}$H twins seems mandatory \cite{Ahn01}. However, the $^3_{\Lambda}$H+$^4_{\Lambda}$H+t mass lies above the initial mass $m_0= m(\Xi^-)+m(^9Be)$ and can therefore not be produced in the $\Xi^-+^9$Be compound production scheme (see right part of Fig.~\ref{fig:bormio04}).
With an energy of 12.6\,MeV below $m_0$, the channel $^4_{\Lambda}$H+$^6_{\Lambda}$He is the most likely twin in the present scenario, followed by the $^4_{\Lambda}$H+$^5_{\Lambda}$He+n decay. Given however the experimental precision of about 1\,\mevc1 for the momentum calibration in E906 \cite{Ahn01}, neither the decays of $^{4}_{\Lambda}$H+$^6_{\Lambda}$He with (133,108){\mevc1} nor decays of
$^{4}_{\Lambda}$H+$^5_{\Lambda}$He pairs with (133,99){\mevc1} seem to explain the structure around (133,114){\mevc1}. Of course in particular the first one could contribute to this enhancement in the tail region. Even more intriguing is however the fact that the statistical models predict a production of $^4_{\Lambda}$H+$^4_{\Lambda}$H twins exceeding the $^3_{\Lambda}$H+$^4_{\Lambda}$H production. Considering furthermore the branching ratios for two-body $\pi^-$ decays of $\Gamma_{\pi^-+^3He}/\Gamma_{total} \approx 0.26$ \cite{Kam98} and
$\Gamma_{\pi^-+^4He}/\Gamma_{total} \approx 0.5$ \cite{Out98} the absence of a bump which could be attributed to $^4_{\Lambda}$H+$^4_{\Lambda}$H is particularly puzzling.
A similar conclusion is obtained \cite{sanchez10} within the quasi-free/rescattering picture of Yamamoto {\em et al.} \cite{Yam97b} resulting in the production of excited  $^8_{\Lambda\Lambda}$He or  $^8_{\Lambda\Lambda}$H nuclei.

At first sight it seems that an alternative production process than the ones discussed so far is required to explain the singular structure at (133,114){\mevc1} in terms of $^3_{\Lambda}$H+$^4_{\Lambda}$H twins.
Note however that in the initial analysis of the E906 data the bump at (133,114){\mevc1} served as a calibration point for the pion momenta, taking the decay of $^3_{\Lambda}$H+$^4_{\Lambda}$H twins as granted  \cite{Ahn01,thesis}. If that structure were indeed caused by $^{4}_{\Lambda}$H+$^6_{\Lambda}$He twins with (133,108){\mevc1}, it would of course influence the momentum scale in the region of the (114,104){\mevc1} bump. This bump would then be shifted to approximately (108,97){\mevc1}. Considering the uncertainty of ${\Delta}B_{\Lambda\Lambda}$ also such a momentum scale would be compatible with the decay of $^7_{\Lambda\Lambda}$He and $^8_{\Lambda\Lambda}$Li or a mixture of both. The absence of $^9_{\Lambda\Lambda}$Li nuclei may be related to the decreasing pionic decay probability with increasing nuclear mass. Clearly, the present statistical decay model needs to be complemented by quantitative weak decay calculations (see e.g. Ref.~\cite{Gal}) to further corroborate our conjecture.

\section{Antihyperons in Nuclei}

\label{intro}
Concerning antibaryons, realiable information on their nuclear potential are available only
for antiprotons.
Antihyperons annihilate quickly in normal nuclei and spectroscopic
information is therefore not directly accessible.
Mishustin recently suggested to study deeply bound
antibaryonic nuclei via various characteristic signals in their
decay process~\cite{Mish05,Lar08}. It is however not obvious whether
these proposed observables will provide unique and quantitative
signals of deeply bound antihyperonic systems.

Quantitative information on the antihyperon potentials relative to
that of the corresponding hyperon may be obtained via exclusive
antihyper\-on-hyperon pair production close to threshold in
antipro\-ton- nucleus interactions \cite{Poc08}. Once these hyperons
leave the nucleus and are detected, their asymptotic momentum
distributions will reflect the depth of the respective potentials.
However, since in the {\Pap\Pp} center-of-mass the distribution of
the produced baryon-antibaryon pair will usually not be isotropic,
the analysis can rely only on the {\em transverse} momenta of the
outgoing baryons. In Ref.~\cite{Poc08} it was demonstrated that the
individual transverse momentum distributions alone do not allow to
extract unambiguous information on the potential of antihyperons.
Instead, the transverse momentum asymmetry $\alpha_{T}$ defined in
terms of the transverse momenta of the {\em coincident} particles
can be used to explore event-by-event correlations:
\begin{equation}
\alpha_{T}=\frac{p_{T}(\PgL)-p_{T}(\PagL)}{p_{T}(\PgL)+p_{T}(\PagL)}.
\label{eq:00}
\end{equation}
In a purely classical, non-relativistic picture this asymmetry is of
the order of ${\Delta}U/4\cdot E_0$, where ${\Delta}U$ is the
potential difference and $E_0$ the typical kinetic energy of the
hadrons.

\subsection{Antihyperon-Hyperon Pair Production and Propagation in Nuclei}

In Ref. \cite{Poc08} we have examined the influence of the potentials on the
transverse momentum asymmetry of coincident hyperons and
antihyperons by means of a schematic Monte Carlo simulation. Albeit
crude, this classical approach allows to explore the role of
different features of the reaction in a transparent way.

The absorption of the antiprotons entering the target nucleus
determines the points of annihilation inside the nucleus and the
paths which the eventually produced hyperons and antihyperons have
to pass inside the nucleus prior to emission. Because
of the strong absorption of the antihyperons, the {\em emitted}
hyperon-antihyperon pairs are - unlike in inclusive reactions
\cite{Sibirtsev99,Lenske05} - created close to the corona of the
target nucleus at a typical density of 20 to 25\% of the central
nuclear density. In reactions close to threshold the Fermi motion of the protons
inside the nuclear target contributes significantly to the final
momenta. Lacking any detailed experimental
information it is assumed that the annihilation cross sections for
antihyperons show a similar momentum dependence as the \Pap\Pp\
system \cite{Ben94,Weber02}.

The energy and the momentum of the baryons propagating within the
nucleus are related according to \cite{Yamazaki99}:
\begin{equation}
(E-V)^2=(M_0+S)^2+\mathbf{P_{in}}^2 \label{eq:05}
\end{equation}
Here $V$ and $S$ denote the real part of the vector and scalar
potential, respectively. The relation between the momenta inside and
outside of the nuclear potential are approximated by
\begin{equation}
\mathbf{P_{out}}^2 + M_0^2=(\sqrt{(M_0+S)^2+\mathbf{P_{in}}^2}+V)^2.
\label{eq:06}
\end{equation}
Refractive effects at the potential boundary were ignored.
The default parameters for the scalar and vector potentials of the
various baryons at normal nuclear density $\rho_0$ were adopted
from Ref.~\cite{Sai05}.
Since the antiproton
annihilation and the subsequent antihadron-hadron pair production take place
in the nuclear periphery at low densities $\rho$ we assumed a linear
density dependence $\propto \rho/\rho_0$ for all vector and scalar
potentials.

\begin{figure}[t]
\begin{center}
\includegraphics[width=0.90\linewidth]{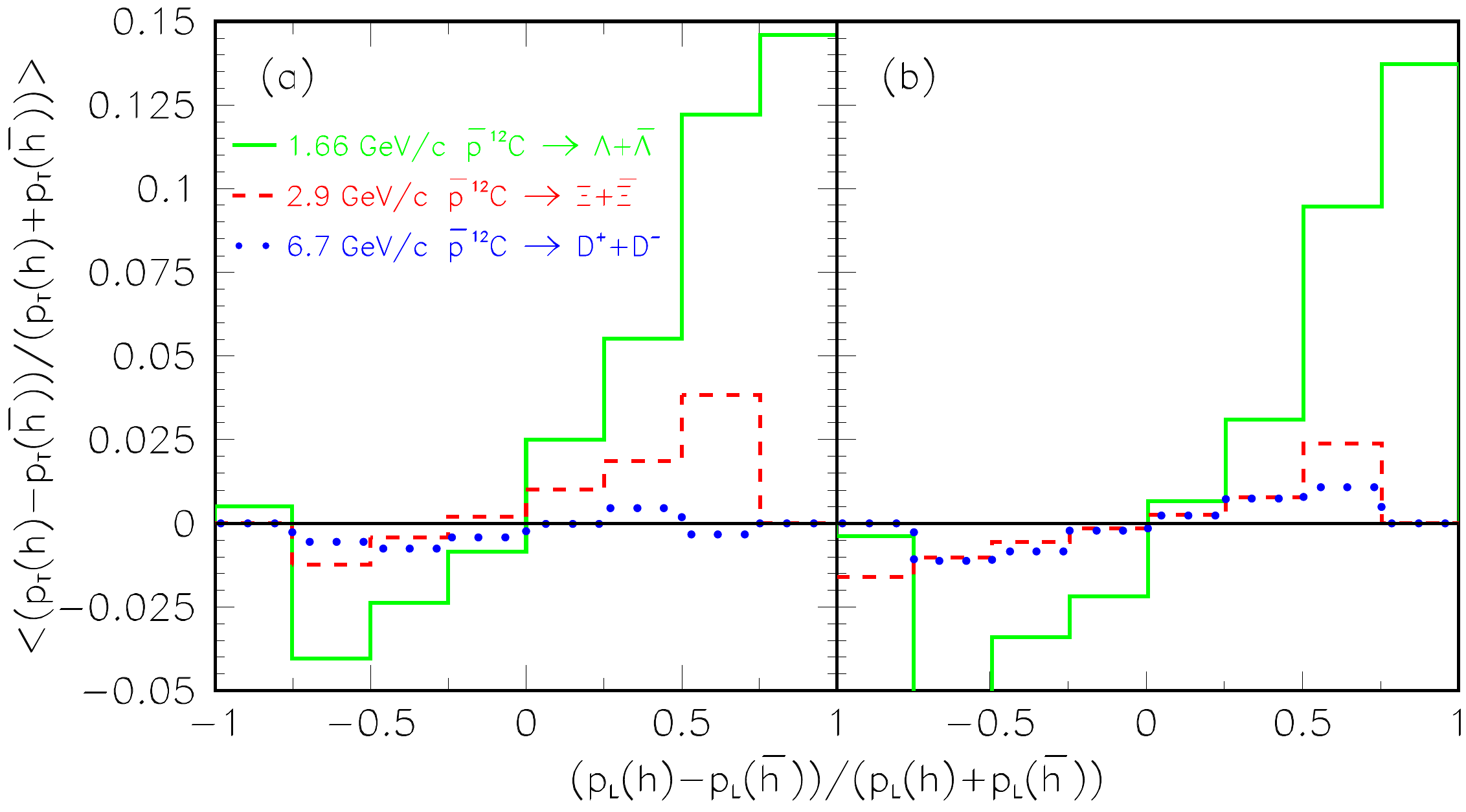}
\end{center}
\caption{Left part: Average transverse momentum asymmetry of
{\PgL\PagL}(solid line), \PgX\PagX\ (dashed line) and D$^+$D$^-$
pairs (dotted line) produced exclusively in 1.66, 2.9 and 6.7\gevc1\
\Pap $^{12}$C interactions, respectively. Right part: the same as in
part (a) but taking the momentum dependence of the potential into account.}
\label{fig:bormio05}
\end{figure}

\subsection{Transverse Momentum Correlations at $\overline{\rm P}$ANDA}

For a compact representation of event-by-event correlations we
examine the transverse momentum asymmetry $\alpha_{T}$ as a function
of the longitudinal asymmetry $\alpha_{L}$, where
 $\alpha_{L}$ is defined by analogy to Eq.~\ref{eq:00} in terms of the
longitudinal momentum components:
\begin{equation}
\noindent
\alpha_{L}=\frac{p_{L}(\PgL)-p_{L}(\PagL)}{p_{L}(\PgL)+p_{L}(\PagL)}
.~~ \label{eq:07}
\end{equation}
In future more sophisticated analysis methods which e.g. use mixed
events as a reference, further details of these
correlations may be revealed.

Transverse momentum correlations can in principle be analyzed for
each hadron-antihadron pair produced exclusively in $\overline{p}A$
interactions. As examples, the solid, dashed and dotted
histograms in figure \ref{fig:bormio05}a show the average transverse
momentum asymmetries of {\PgL\PagL}, {\PgX\PagX} and D$^+$D$^-$
pairs produced in 1.66, 2.9 and 6.7\gevc1\ \Pap\ + $^{12}$C
interactions, respectively. For simplicity, isotropic center-of-mass
distributions were assumed in case of the {\PgX\PagX} and D$^+$D$^-$
production. For the \PgX\ and \PagX\ baryons the same absorption
cross sections as for the \PgL\ and \PagL\ were adopted, whereas for
D$^-$ and D$^+$ mesons energy independent absorption cross sections
of 10 and 90 mb, respectively, were taken.
In the left part of Fig.~\ref{fig:bormio05} no momentum dependence
of the potentials was considered.

In line with the classical picture mentioned above ($\alpha_T
\propto {\Delta}U/4\cdot E_0$) the smaller asymmetries for the
heavier particles are a consequence of the the large \PgX\ hyperon
and D meson laboratory momenta. In the simulations of
Fig.~\ref{fig:bormio05}b the momentum dependence of the potentials was
added by means of the scaling factor \cite{Gale87}.
As expected the transverse asymmetry is shifted
towards more negative values. Nevertheless the asymmetry remains sizable
even for the {\PgX\PagX} case.  At {\Panda} when reaching the design luminosity a measurement of
$\alpha_T$ for {\PgX\PagX} pairs with a precision of about 10\%  will require
a measurement time of typically less than a day. Furthermore it puts only moderate constraints
on the detector performance, e.g. the tracking capabilities of the central vertex detector.

The fact that energy and momentum conservation are the main
ingredients of the proposed method raises hope that similar results
might be obtained by more sophisticated calculations. Since most of
the {\em emitted} hyperon-antihyperon pairs are created in the
nuclear periphery at subsaturation density, a neutron skin of
neutron rich target nuclei may help to explore different effective
potentials. Possible deflections at the potential boundary which are
ignored in the present work may be at least partly eliminated by
demanding that the target nucleus remains intact. Furthermore, it
may be interesting to study questions related to the formation
time by using target nuclei of different size. Finally,
it should be noted that this method can be extended to the case of
photo- or even electro-production of short-lived resonances in
nuclei decaying into particle-antiparticle pairs.

This work was supported by the BMBF under Contract numbers 06MZ225I and 06MZ9182. A.S.L. acknowledges the support from the State of Rhineland-Palatinate via the Research Centre `Elementary Forces and Mathematical Foundations' (EMG). We also thank the European Community-Research Infrastructure Integrating Activity {\em Study of Strongly Interacting Matter} (HadronPhysics2, Grant Agreement n. 227431; SPHERE network) under the Seventh Framework Programme of EU for their support.


\begin{thebibliography}{99}

\bibitem{Lee56}
T.D. Lee and C.N. Yang, Nuovo Cim. {\bf 3}, 749 (1956).

\bibitem{Duer56a}
H.-P. D\"urr and E. Teller, Phys. Rev. {\bf 101}, 494
(1956); H.-P. D\"urr, Phys. Rev. {\bf 103}, 469 (1956).

\bibitem{Duer58}
Hans-Peter D\"urr, Phys. Rev. {\bf 109}, 1347 (1958).

\bibitem{Gol58b}
G. Goldhaber and J. Sandweiss, Phys. Rev. {\bf 110}, 1476 (1958).

\bibitem{Ben94}
G. Bendiscioli and D. Kharzeev,
Rivista del Nuovo Cimento {\bf 17}, 1 (1994).

\bibitem{Bar72}
P. D. Barnes {\it et al.}, Phys. Rev. Lett. {\bf 29}, 1132 (1972).

\bibitem{Bac72}
G. Backenstoss {\it et al.}, Physics Letters B {\bf 41}, 552 (1972).

\bibitem{Rob77}
P. Roberson {\it et al.}, Phys. Rev C {\bf 16}, 1945 (1977).

\bibitem{Pot78}
H. Poth {\it et al.}, Nucl. Phys. A {\bf 294}, 435 (1978).

\bibitem{Fri07}
E. Friedman and A. Gal, Phys. Rep. {\bf 452}, 89, (2007).

\bibitem{Aue81}
E. H. Auerbach, C. B. Dover, and S. H. Kahana, Phys. Rev. Lett. {\bf
46}, 702 (1981).

\bibitem{Bat81}
C.J. Batty, Nucl. Phys. A {\bf 372}, 433 (1981).

\bibitem{Fri05}
E. Friedman, A. Gal and  J. Mares, Nucl. Phys. A {\bf 761}, 283
(2005).

\bibitem{Teis94}
Stefan Teis {\it et al.}, Phys. Rev. C {\bf 50}, 388 (1994).

\bibitem{Spie96}
C. Spieles {\it et al.}, Phys. Rev. C {\bf 53}, 2011 (1996).

\bibitem{Sib98}
A. Sibirtsev {\it et al.}, Nucl. Phys. A {\bf 632}, 131 (1998).

\bibitem{Dov80}
C.B. Dover and J.M. Richard, Phys. Rev. C {\bf 21}, 1466 (1980).

\bibitem{Fae82}
A. Faessler, G. L\"ubeck and K. Shimizu, Phys. Rev. D {\bf 26}, 3280
(1982).

\bibitem{Ker73}
A.K.Kerman and M.S. Weiss,
Phys. Rev. C {\bf 8}, 408 (1973).

\bibitem{Bot07} A.S. Botvina and J. Pochodzalla,
Phys. Rev. C {\bf 76}, 024909 (2007).

\bibitem{STAR}
The STAR Collaboration, Science  328, 58-62 (2010).

\bibitem{Yam94}
Y. Yamamoto, M. Sano and M. Wakai,
Prog. Theor. Phys. Suppl. {\bf 117}, 265 (1994).

\bibitem{Yam97}
T. Yamada and K. Ikeda,
Phys. Rev. C {\bf 56}, 3216 (1997).

\bibitem{Yam97b}
Y. Yamamoto, M. Wakai, T. Motoba, T. Fukuda,
Nucl. Phys. A {\bf 625}, 107 (1997).

\bibitem{Hir97}
Y. Hirata {\em et al.},
Nucl. Phys. A {\bf 639}, 389c (1998).

\bibitem{Hir99} Y. Hirata {\em et al.},
Prog. Theor. Phys. {\bf 102}, 89 (1999).

\bibitem{Aok09}
S. Aoki {\em et al.}, KEK E176 Collaboration,
Nucl. Phys. A {\bf 828}, 191 (2009).

\bibitem{Ohn01}
A. Ohnishi, Y. Hirata, Y. Nara, S. Shinmura and Y. Akaishi,
Nucl. Phys. A {\bf 684}, 595 (2001).

\bibitem{Ike94}
K. Ikeda {\em et al.},
Prog. Theor. Phys. {\bf 91}, 747 (1994).

\bibitem{Yam94b}
Y. Yamamoto {\em et al.},
Prog. Theor. Phys. Suppl. {\bf 117}, 281 (1994).

\bibitem{Bat99}
C.J. Batty, E. Friedman, A. Gal,
Phys. Rev. C {\bf 59}, 295-304(1999).

\bibitem{Poc04}
J. Pochodzalla,
Nucl. Instr. Meth B {\bf 214}, 149 (2004).

\bibitem{PandaTPR}
\Panda\ Collaboration, Technical Progress Report
(GSI Darmstadt), pp. 1-383 (2005).

\bibitem{PandaPhysicsbook}
\Panda\ Collaboration, Physics Performance Report
for PANDA,
arXiv:0903.3905.

\bibitem{sanchez10}
A. Sanchez Lorente, A. Botvina, J. Pochodzalla, submitted for publication.

\bibitem{Fermi}
E. Fermi,
Progr. Theor. Phys. {\bf 5}, 570 (1950).

\bibitem{Dov83}
C.B. Dover and A. Gal, Ann. Phys. {\bf 147}, 309 (1983).

\bibitem{Aok95}
S. Aoki {\em et al.}, Phys. Lett. B {\bf 355}, 45 (1995).

\bibitem{Wil59}
D.H. Wilkinson, S.J. St. Lorant, D.K. Robinson, and S. Lokanathan,
Phys. Rev. Lett. {\bf 3}, 397 (1959).

\bibitem{Ste63}
P.H. Steinberg and R.J. Prem,
Phys. Rev. Lett. {\bf 11}, 429 (1963).

\bibitem{Bec68}
A. Bechdolff, G. Baumann, J.P. Gerber, and P. C\"uer,
Phys. Lett. {\bf 26B}, 174 (1968).

\bibitem{Aok93}
S. Aoki {\em et al.}, Prog. Theor. Phys. {\bf 89}, 493 (1993).

\bibitem{Ich01}
A. Ichikawa {\em et al.}, Phys. Lett. B {\bf 500}, 37 (2001).

\bibitem{Ahn01}
J.K. Ahn {\em et al.},
Phys. Rev. Lett. {\bf 87}, 132504-1 (2001).

\bibitem{Kum02}
Izumi Kumagai-Fuse and Shigeto Okabe,
Phys. Rev. C {\bf 66}, 014003 (2002).

\bibitem{Ran07}
S.D. Randeniya and E.V. Hungerford,
Phys. Rev. C {\bf 76}, 064308 (2007).


\bibitem{Kam98}
H. Kamada,J. Golak, K. Miyagawa, H. Witala, W. Gl\"ockle,
Phys. Rev. C {\bf 57}, 1595 (1998).

\bibitem{Out98}
H. Outa {\em et al.}, Nucl. Phys. A{\bf 639}, 251c (1998).

\bibitem{thesis}
Joe P. Nakano, {\em Study of Double-Lambda Hypernuclei by Using Cylindrical Detector System}, Ph.D. Thesis, Center for Nuclear Study, University of Tokyo, Report CNS-REP-28 (2000).

\bibitem{Gal}
A. Gal, Nucl.Phys.A {\bf 828}, 72 (2009).

\bibitem{Mish05}
I.N. Mishustin {\it et al.}, Phys. Rev. C {\bf 71}, 035201 (2005).

\bibitem{Lar08}
A.B. Larionov {\it et al.}, Phys.Rev.C{\bf 78}, 014604 (2008).

\bibitem{Poc08}
J. Pochodzalla, Phys. Lett. B {\bf 669}, 306 (2008).

\bibitem{Sibirtsev99}
A. Sibirtsev, K. Tsushima, and A.W. Thomas, Eur. Phys.J. A{\bf 6},
351 (1999).

\bibitem{Lenske05}
H. Lenske and P. Kienle, Phys. Lett. B {\bf 647}, 82 (2007).

\bibitem{Weber02}
H. Weber, E.L. Bratkovskaya, and H. St\"ocker, Phys. Rev. C {\bf
66}, 054903 (2002).

\bibitem{Yamazaki99}
T. Yamazaki and Y. Akaishi, Phys. Lett. B {\bf  453}, 1 (1999).

\bibitem{Sai05}
K. Saito, K. Tsushima, and A.W. Thomas, Prog.Part.Nucl.Phys. {\bf
58}, 1 (2007).

\bibitem{Gale87}
C. Gale, G. Bertsch and S. Das Gupta, Phys. Rev C {\bf 35}, 1666
(1987).


\end{thebibliography}
\end{document}